\documentclass[
superscriptaddress,
reprint,
 amsmath,amssymb,
 aps, prl,
 floatfix,
]{revtex4-2}
\usepackage{graphicx}
\usepackage[cmyk]{xcolor}
\usepackage{overpic}
\usepackage{dcolumn}
\usepackage{bm}
\usepackage{microtype}
\usepackage{hyperref}

\begin{document}

\preprint{APS/123-QED}

\title{\textbf{Generation of period-tunable MeV few-attosecond electron pulse trains via counter-propagating lasers} 
}%

\author{Qi Huang}
 \affiliation{%
Department of Plasma Physics and Fusion Engineering, University of Science and Technology of China, Hefei, Anhui 230026, People’s Republic of China.
}
\author{Qing Jia}%
 \email{qjia@ustc.edu.cn}
\affiliation{%
Department of Plasma Physics and Fusion Engineering, University of Science and Technology of China, Hefei, Anhui 230026, People’s Republic of China.
}%

\author{Zhongxuan Wang}
 \affiliation{%
Department of Plasma Physics and Fusion Engineering, University of Science and Technology of China, Hefei, Anhui 230026, People’s Republic of China.
}

\author{Jian Zheng}
 \affiliation{%
Department of Plasma Physics and Fusion Engineering, University of Science and Technology of China, Hefei, Anhui 230026, People’s Republic of China.
}
\affiliation{
Collaborative Innovation Center of IFSA, Shanghai Jiao Tong University, Shanghai 200240, People’s Republic of China.
}


\date{\today}

\begin{abstract}

Attosecond electron pulses permit real-time probing of ultrafast material dynamics. However, generating few-attosecond electron pulses with MeV energies and low energy spread remains an enduring challenge for conventional beam-modulation techniques. Here we propose a compact dual-laser scheme to modulate readily accessible electron beams into few-attosecond pulse trains, leveraging a stable parametric-resonance regime coupled with direct laser acceleration. An accompanying theoretical framework is developed, yielding closed-form expressions for the tunable pulse period, duration, energy modulation and formation time, enabling flexible customization of the produced attosecond pulse trains. Consistent with these theoretical predictions, simulations verify the generation of $\sim$ 1 as pulses with a Lorentz factor up to 15 and a relative energy spread below 0.02\%. This work offers an experimentally feasible pathway toward high-quality, tunable MeV few-attosecond electron pulses. 


\end{abstract}

\maketitle

Attosecond electron pulses have emerged as a powerful tool to resolve atomic-scale light-matter interactions \cite{García2025,Gaida2024,Yong2022}, with applications including ultrafast electron diffraction \cite{Dandan2024,Ahmed2010} and microscopy \cite{Priebe2017,morimoto2022}.
Going to even shorter pulse durations—few-attosecond or even zeptosecond—would allow us to penetrate beyond the atomic scale to resolve nuclear-level processes, opening new frontiers in nuclear physics\cite{Rietz2011,Chiara2018,Vanacore_2018}.

For such applications, electron beams with kinetic energies from tens of keV to MeV (Lorentz factor $\gamma\sim 1$–10) and a relative energy spread below 1\% are typically required \cite{filippetto2022}. 
While 100‑keV beams have been modulated into $100$-as pulses via laser‑dielectric or ponderomotive interactions \cite{peter2007,Hommelhoff2018}, MeV beams have reached few-femtosecond to sub‑femtosecond bunching through radiofrequency (RF) or THz modulation \cite{Schaap2025,Zhao2018}. All these methods rely on imprinting an energy modulation $\Delta W$ that compresses the beam during drift \cite{Nabben2023,Kozák2018}.
Liouville’s theorem sets a minimum pulse duration $\tau_{\min}\propto \Delta W_0/\Delta W$, where $\Delta W_0$ is the uncorrelated energy spread of the initial beam. The longitudinal focal length $L_f$---the drift distance over which the attosecond structure forms and then broadens---scales with $\tau_{\min}$ and $\gamma$ as $L_f \propto \gamma^2 \sqrt{\gamma^2-1}/\Delta W \propto \gamma^2 \sqrt{\gamma^2-1}\,\tau_{\min}/\Delta W_0$ \cite{Norbert2019}. Under typical 100‑keV ultrafast microscopy conditions ($\tau_{\min}\approx 100$ as, $\gamma\sim 1.2$, $\Delta W_0\sim 1$ eV) \cite{Morimoto2018,Black2019}, $L_f$ is on the order of millimeters, giving researchers sufficient length for further experiments and measurements.

A severe challenge emerges, however, when one pushes $\tau_{\min}$ into the few‑attosecond or sub‑attosecond regime. For 100‑keV beams, the scaling relation forces $L_f$ down to the micrometer scale, leaving insufficient length for diagnosis and application. Extending $L_f$ therefore requires boosting the electron energy to the MeV level. Although conventional RF accelerators can deliver high-$\gamma$ beams, their uncorrelated energy spread $\Delta W_0$ is typically on the keV level \cite{Weathersby2015,Schaap2025}. To reach sub‑attosecond pulse durations, $\Delta W$ would need to be in the MeV range---far beyond the tolerance of typical diffraction applications. Generating high‑quality MeV, attosecond‑to‑sub‑attosecond electron pulses therefore remains an elusive goal. 

\begin{figure} 
	\centering
	\begin{overpic}[width=0.48\textwidth]{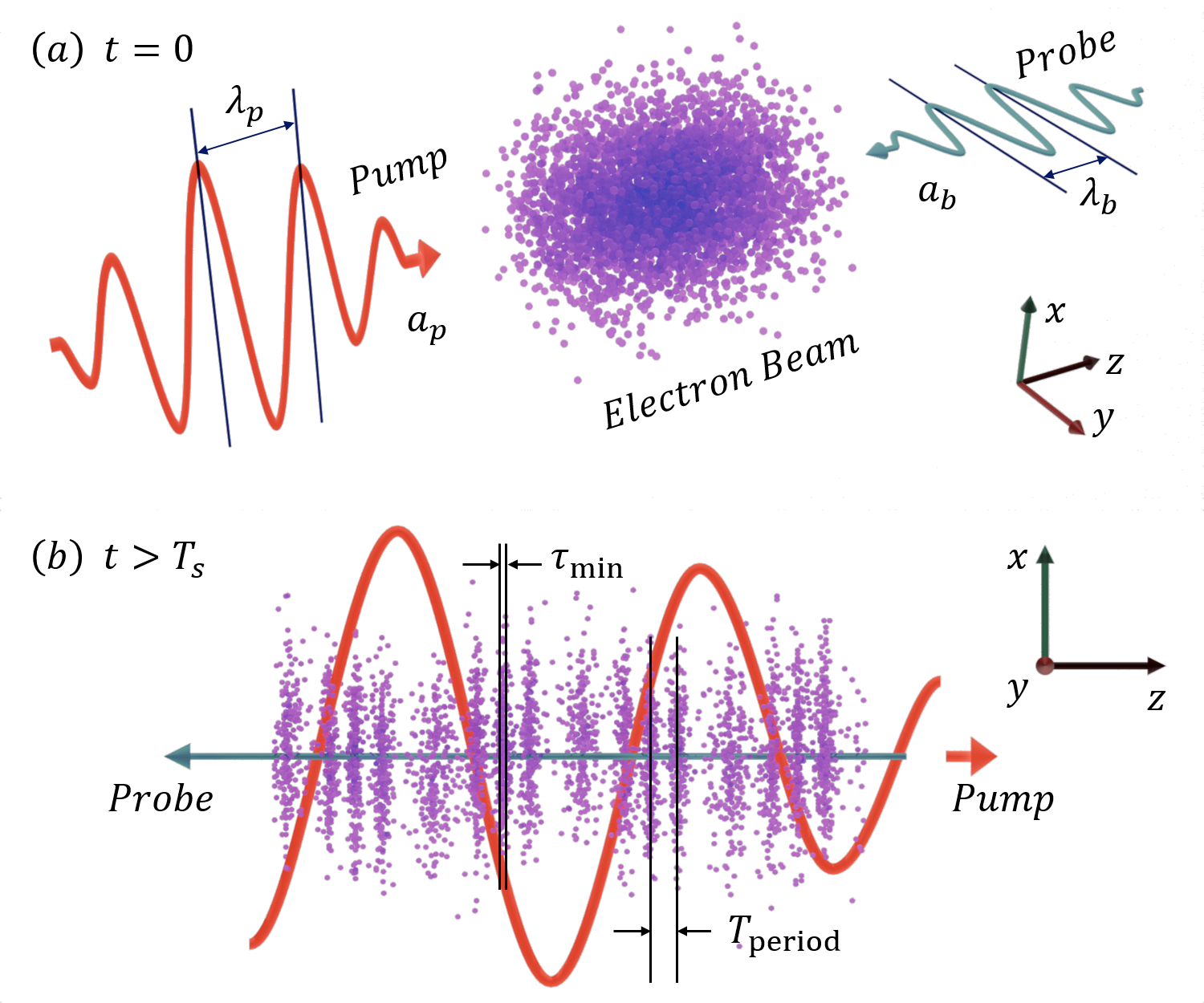}
        \end{overpic}

	\caption{Schematic of attosecond electron pulse train generation using two counter-propagating lasers.
    (a) The $x$-polarized pump laser ($a_p$) is used for DLA, while the probe laser ($a_b$) provides subsequent beam modulation.
    (b) When $t>T_s$, the electron beam is modulated into pulse trains with minimum duration $\tau_{\min}$ and tunable period $T_{\text{period}}$.
}
	\label{fig:LightScenario} 
\end{figure}

In this work, we address this challenge using a compact dual-laser scheme. As illustrated in Fig. \ref{fig:LightScenario}(a), a relativistic-intensity pump pulse and a counter-propagating weak probe pulse interact with an electron beam of small $\Delta W_0$. After the characteristic formation time $T_s$, the electron beam is modulated into high-quality MeV few-attosecond pulse trains with minimum duration $\tau_{\min}$ and tunable period $T_{\text{period}}$, as shown in Fig. \ref{fig:LightScenario}(b). This modulation arises from a previously unexplored stable regime that combines direct laser acceleration (DLA) \cite{arefiev2015} with parametric resonance \cite{fossen2011}. Distinct from conventional plasma-based DLA schemes \cite{Sun2024,Shi2021,Wang2019}, which produce $\sim\!100$-as pulse trains and offer no active control over $\tau_{\min}$ or $\Delta W$, this method enables flexible customization of the resulting few-attosecond to sub-attosecond pulse trains.


Let us first analyze the modulation of an electron initially at rest by two orthogonally polarized, counter-propagating lasers. The pump laser is $x$-polarized with normalized vector potential $a_p=a_{p0}\sin(k_p\xi)$ for initial acceleration, and the probe laser is $y$-polarized with $a_b=a_{b0}\sin(k_b\mu)$ for subsequent modulation, where $\xi=z-ct$, $\mu=z+ct$, and $a_{b0}\ll a_{p0}$ ($c$ is the speed of light). 
For the sub‑pC electron beam and picosecond interaction timescale considered here, space‑charge effects are negligible \cite{filippetto2022}. The momentum equation can therefore be obtained as:
\begin{equation}
\frac{d}{dt}\boldsymbol{P} = -\frac{e}{m c}\left[\boldsymbol{E}_p+\boldsymbol{E}_b + c\boldsymbol{\beta}\times(\boldsymbol{B}_p+\boldsymbol{B}_b)\right],
\label{eq:motion_Equation}
\end{equation}
where $e$ and $m$ are the elementary charge and electron mass, and $\boldsymbol{P}=\gamma\boldsymbol{\beta}$ is the normalized momentum ($\boldsymbol{v}=c\boldsymbol{\beta}$). The electromagnetic fields are given by $\boldsymbol{E}_i = -(mc/e)\partial_t \boldsymbol{a}_i$ and $\boldsymbol{B}_i = (mc/e)\nabla\times \boldsymbol{a}_i$, where $i=p,\,b$. 
We then expand the electron velocity and momentum to second order: $\boldsymbol{\beta}=\boldsymbol{\beta}_p+\boldsymbol{\beta}_1+\boldsymbol{\beta}_2$, $\boldsymbol{P}=\boldsymbol{P}_p+\boldsymbol{P}_1+\boldsymbol{P}_2$. With initial conditions ${\beta}_{1x}={\beta}_{1z}=0$ at $t=0$ and defining $\gamma_p=\sqrt{1+P_p^2}$, the hierarchical equations are derived as follows:
\begin{equation}
P_{px}=a_p,P_{pz}=a_p^2/2,P_{py}=0
    \label{eq:hierarch1}
\end{equation}
\begin{equation}
P_{1y}=\gamma_p \beta_{1y}=a_b, P_{1x}=P_{1z}=0
    \label{eq:hierarch2}
\end{equation}
\begin{equation}
d_t P_{2z}=-\beta_{1y}\partial_z a_b, P_{2x}=P_{2y}=0
    \label{eq:hierarch3}
\end{equation}
Eq.(\ref{eq:hierarch1}) describes the well-established DLA of electrons in vacuum. For an electron initially at $z(t=0)=\mu_0'$, its z-direction trajectory evolves as $z_p=-a_{p0}^2[\xi+\sin(2k_p \xi)/(2 k_p)]/4+\mu_0'$, exhibiting acceleration-deceleration cycles. 

From Eq.~(\ref{eq:hierarch2}) and Eq.~(\ref{eq:hierarch3}), the $z$-direction perturbation $\delta z = z - z_p$ satisfies the following Hill‑type equation \cite{fossen2011}:
\begin{equation}
\frac{d^2}{d\xi^2} \delta z + \kappa(\xi) \delta z = a_{b0}^2 k_b (d_\xi \mu_p + 2 d_\xi \delta z) \sin(2k_b \mu_p),
\label{eq:hill_Equation}
\end{equation}
where $\kappa(\xi) = -4 a_{b0}^2 k_b^2 (d_\xi \mu_p) \cos(2k_b \mu_p)$ and $\mu_p = 2z_p - \xi$. The detailed derivation is given in Appendix~A.

Equation~(\ref{eq:hill_Equation}) describes a parametrically driven longitudinal motion of the electron. Physically, the pump laser drives a periodic acceleration‑deceleration motion, while the probe field modulates this motion. Strong amplification of $\delta z$ occurs when the electron experiences an integer number of probe cycles per acceleration–deceleration period, thus sampling the same phase of the probe field and enabling a resonant modulation. The corresponding resonance condition can be written as:
\begin{equation}
\left(1 + \frac{a_{p0}^2}{2}\right) \frac{k_b}{k_p} = n \in \mathbb{N},
\label{eq:condition_and_timescale1}
\end{equation}
When this condition is satisfied, the longitudinal electron positions become strongly perturbed, and after a characteristic time a pronounced density modulation develops along the beam, resulting in $n$ electron pulses within half the pump wavelength.

\begin{figure}
   \centering
    \begin{overpic}[width=0.48\textwidth]{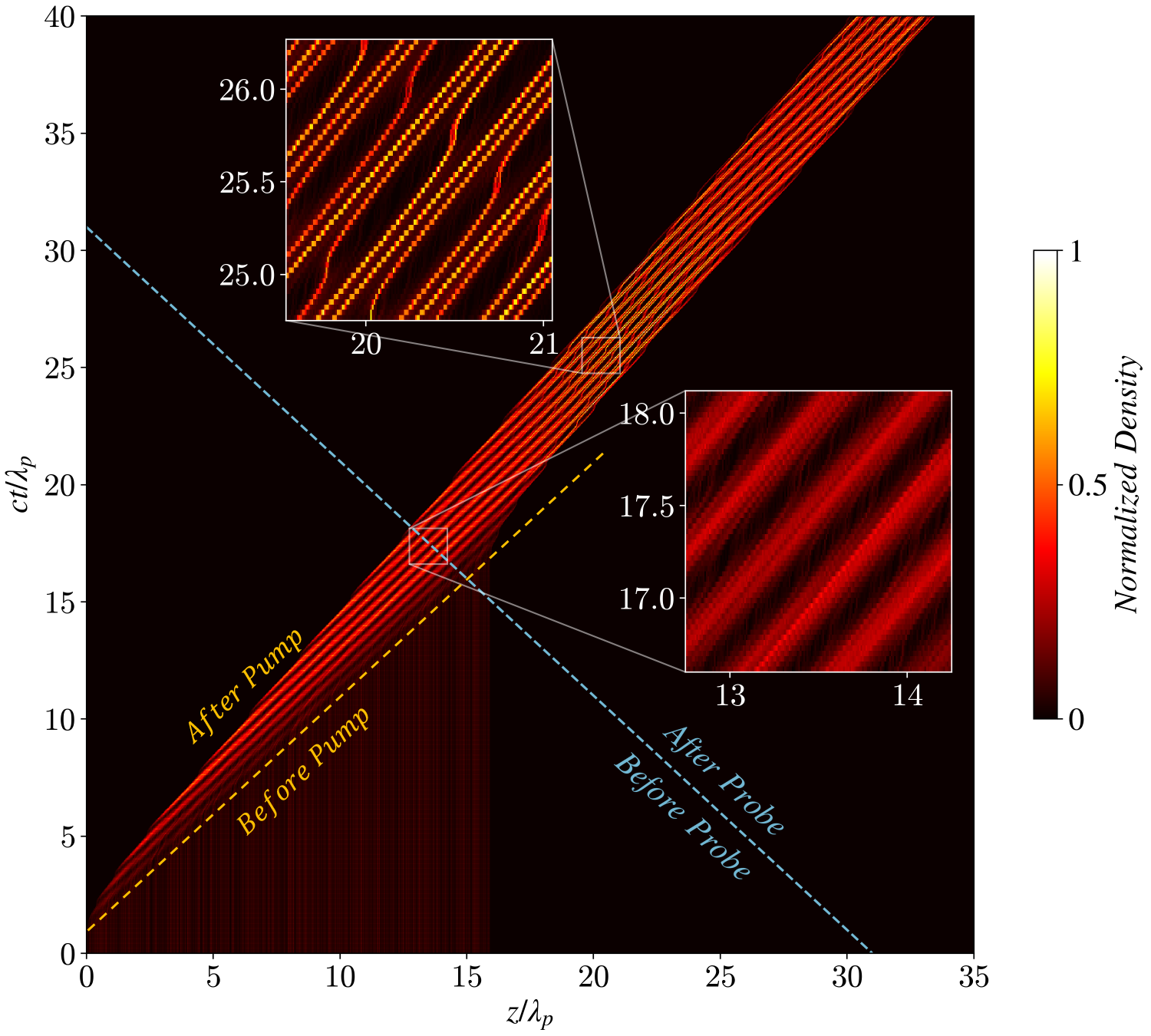}
        \put(10,82){\textcolor{white}{(a)}}
        \put(60,52){\textcolor{white}{(b)}}
        \put(25,82){\textcolor{white}{(c)}}
        \end{overpic}

	\caption{(a) Spatiotemporal evolution of attosecond pulse train formation. The color scale represents the electron density distribution. (b) After the application of the pump laser, the electron beam is modulated into bunches spaced by $\pi/k_p$. (c) With the subsequent modulation of the probe laser, each primary bunch is further subdivided into $n$ distinct pulses. 
}
	\label{fig:Record1} 
\end{figure}

To verify the theoretical predictions and illustrate the modulation mechanism, we perform particle-tracking simulations. The simulation parameters are $a_{p0}=4$, $a_{b0}=0.2$, and $k_b/k_p=1/3$ [corresponding to $n=3$ predicted with Eq. (\ref{eq:condition_and_timescale1})], with electrons uniformly distributed in longitudinal position $z$ at $t=0$. The lasers are orthogonally polarized and both include a ramp-up of 2.5 optical cycles. 

\begin{table*}
\caption{\label{tab:Formulae}
Formulae for the number of pulses within half the pump wavelength $n$, the characteristic time $T_s$, the energy modulation $\Delta W$, the minimum attosecond pulse duration $\tau_{\min}$, and the maximum Lorentz factor $\gamma_{\max}$. Here $[\mathrm{JJ}]_n\equiv J_n-J_{n+1}$ (where $J_n$ represents the Bessel function of the first kind).}
\begin{ruledtabular}
\begin{tabular}{ccc}
 &$\parallel$&$\perp$\\
\hline
$n$& $\left[\frac{1+\beta_0}{1-\beta_0}\left(1+\frac{a_{p0}^2}{2}\right)\frac{k_b}{k_p}-1\right]/2$& $\frac{1+\beta_0}{1-\beta_0}\left(1+\frac{a_{p0}^2}{2}\right)\frac{k_b}{k_p}$\\
$T_s$& $\frac{(a_{p0}^2/4+1)\pi}{4c}\left[a_{p0}a_{b0}k_pk_b\frac{1-\beta_0}{1+\beta_0} n [\mathrm{JJ}]_{n}(\frac{1+\beta_0}{1-\beta_0}\frac{k_b a_{p0}^2}{4k_p})\right]^{-\frac{1}{2}}$& $\frac{(a_{p0}^2/4+1)\pi}{4a_{b0}c}\left[k_pk_b\frac{1-\beta_0}{1+\beta_0} n J_n(\frac{1+\beta_0}{1-\beta_0}\frac{k_b a_{p0}^2}{2k_p})\right]^{-\frac{1}{2}}$\\
$\Delta W$& $4mc^2\sqrt{{a_{p0}a_{b0}}n\frac{k_p}{k_b}[\mathrm{JJ}]_{n}(\frac{1+\beta_0}{1-\beta_0}\frac{k_b a_{p0}^2}{4k_p})}$& $2mc^2a_{b0}\sqrt{n\frac{k_p}{k_b}J_n(\frac{1+\beta_0}{1-\beta_0}\frac{k_b a_{p0}^2}{2k_p})}$\\
$\tau_{\min}$& \multicolumn{2}{c}{${\left(\frac{1}{1-\beta_0}+\frac{1+\beta_0}{1-\beta_0}\frac{a_{p0}^2}{4}\right)\pi\Delta W_{0}}/({nck_p\Delta W}) $} \\
$\gamma_{\max}$& \multicolumn{2}{c}{$\gamma_0 + a_{p0}^2 / [2(\gamma_0 - \sqrt{\gamma_0^2 - 1})]$} \\
\end{tabular}
\end{ruledtabular}
\end{table*}

The simulation results confirm the predicted modulation dynamics. As shown in Fig. \ref{fig:Record1}(b), when only the pump laser is present, the electron beam is modulated into periodically spaced bunches with a full width at half maximum (FWHM) of approximately $1/4$ of the pump period. With the continued influence of the probe laser, the temporal profile of these bunches is further subdivided into $n$ distinct few-attosecond pulses, where $T_{\text{period}}\approx\pi/(nck_p)$ and the individual pulse duration is much shorter than $\pi/(2nck_p)$, as shown in Fig. \ref{fig:Record1}(c).

A similar resonance condition and characteristic timescale can be derived for parallel‑polarized lasers. For an electron beam with initial velocity $c\beta_0$ (corresponding to the Lorentz factor $\gamma_0=1/\sqrt{1-\beta_0^2}$) and $\Delta W_0$, the generalized expressions for $n$, the characteristic time $T_s$, $\Delta W$, $\tau_{\min}$, and the maximum Lorentz factor $\gamma_{\max}$ are summarized in Table~\ref{tab:Formulae} for both the parallel-polarized ($\parallel$) and the orthogonally polarized ($\perp$) configurations; detailed derivations are given in Appendices B and C. 

\begin{figure}[b]
   \centering
    \begin{overpic}[width=0.48\textwidth]{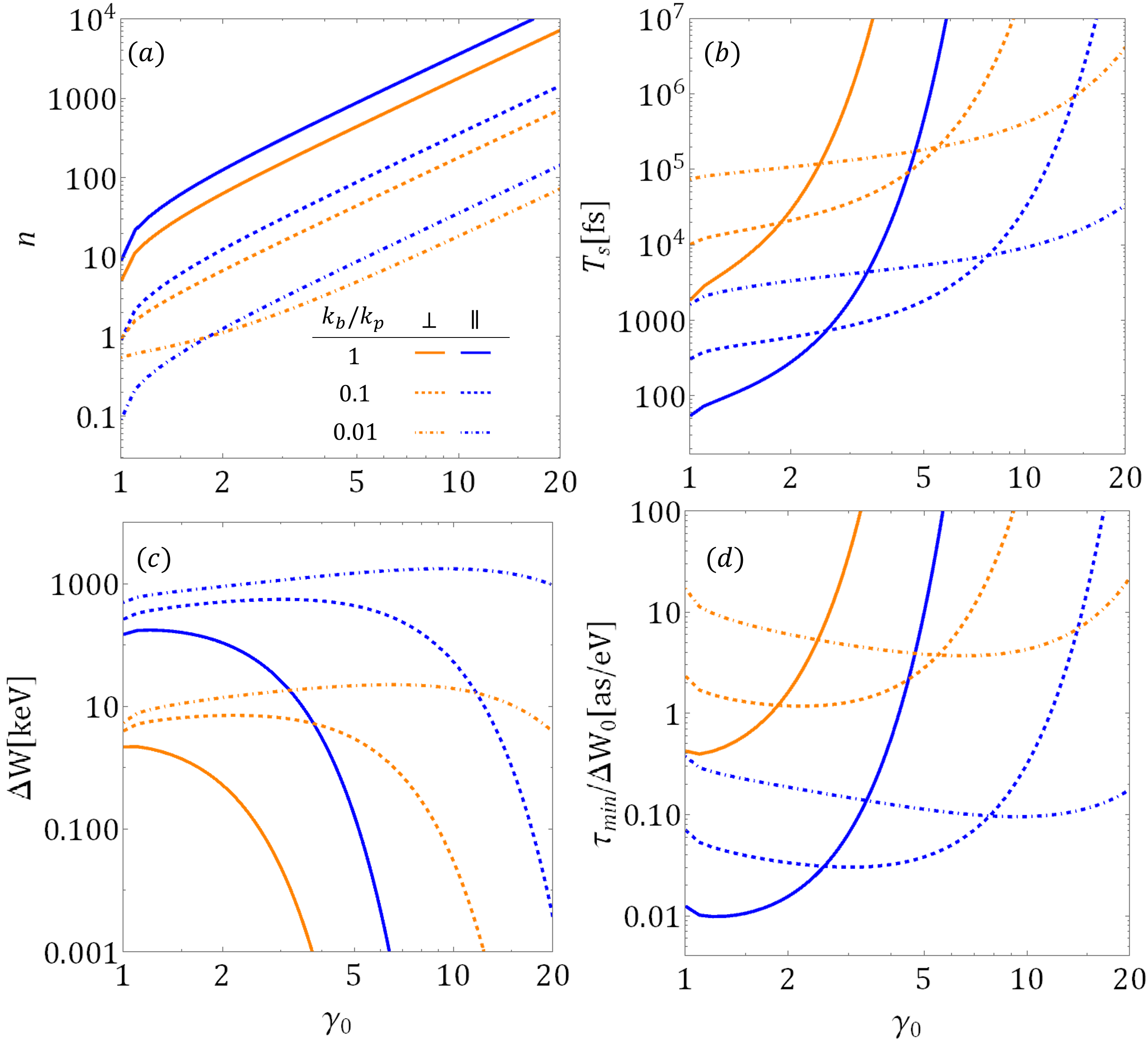}
        \end{overpic}\\

	\caption{Dependence of $n$, $T_s$, $\Delta W$, and $\tau_{\min}$ on $k_b$ and $\gamma_0$ for $a_{p0}=4$, $a_{b0}=0.002$, and $\lambda_p=1$~$\mu$m. Yellow (blue) curves: orthogonal (parallel) polarization. Solid, dashed, and dotted lines: $k_b/k_p=1$, 0.1, and 0.01.
    (a) $n$ increases with both $k_b$ and $\gamma_0$.
(b) Overall, $T_s$ increases with $\gamma_0$ but decreases with $k_b$. For the same parameters, $T_{s\perp}$ is significantly larger than $T_{s\parallel}$.
(c) and (d) Both $\Delta W$ and $\tau_{\min}$ exhibit an extended plateau as $\gamma_0$ varies. A lower probe frequency further extends this plateau region.
 }
	\label{fig:Theory} 
\end{figure}

The parametric dependences predicted by the analytical model are illustrated in Fig.~\ref{fig:Theory}, which shows how $n$, $T_s$, $\Delta W$, and $\tau_{\min}$ vary with the probe wavenumber $k_b$ and the initial Lorentz factor $\gamma_0$. As shown in Fig. \ref{fig:Theory}(a), increasing $\gamma_0$ leads to a larger $n$, indicating that $T_{\text{period}}$ can be tuned by varying the initial kinetic energy of the electron beam. However, an excessively high $\gamma_0$ results in a longer $T_s$ and a smaller $\Delta W$, as seen in Figs. \ref{fig:Theory}(b) and (c), respectively. Since the attosecond pulse duration scales as $\tau_{\min}\propto1/\Delta W$, the reduction in energy modulation broadens the pulse duration, as illustrated in Fig. \ref{fig:Theory}(d). The modulation scheme fails when the pulse duration exceeds the temporal separation between adjacent pulses. Therefore, for high $\gamma_0$, lowering $k_b$ can increase the temporal separation between adjacent pulses while approximately preserving the energy-modulation capability, thereby obtaining viable attosecond electron pulse trains. 
However, if the parametric resonance is too strong, stochastic acceleration occurs. This imposes a threshold on the product of the laser strengths \cite{Sheng2002,zhang2019}, beyond which stable modulation can no longer be maintained.
\begin{figure*}
   \centering
    \begin{overpic}[width=0.95\textwidth]{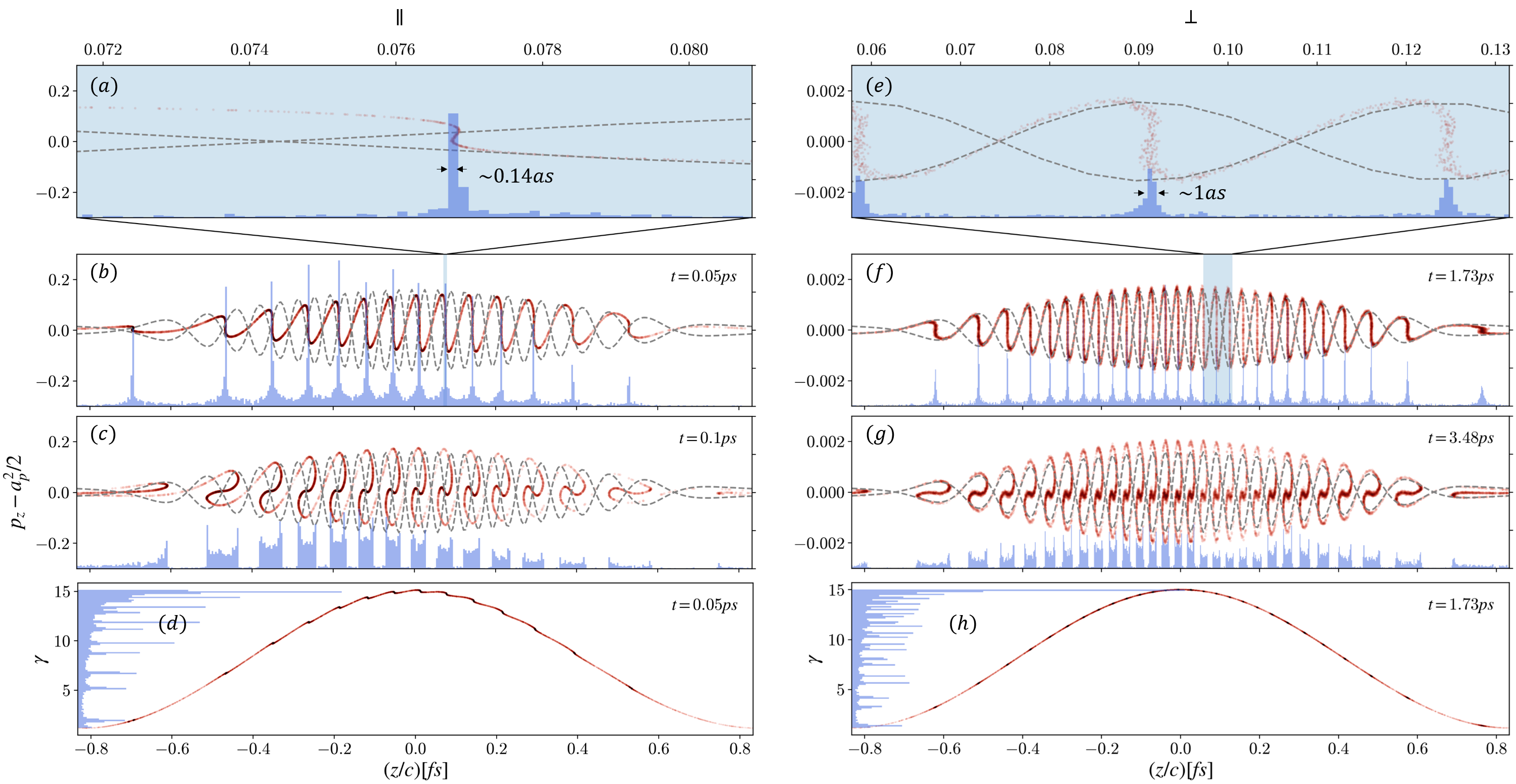}
        \end{overpic}\\

	\caption{Phase-space distributions for $a_{p0}=4$, $a_{b0}=0.002$, $\beta_0=0.5$, $\Delta W_0=4$ eV, and $\lambda_p=\lambda_b=1\ \mu$m. (a)–(c) Parallel polarization configuration; (e)–(g) orthogonal polarization configuration. Blue histograms: longitudinal particle density; gray dashed curves: fitted phase-space boundary. (a),(e) Magnified views of the boxed regions in (b),(f), showing FWHM pulse durations of $0.14$ as and $1$ as, respectively. (d),(h) Corresponding energy spectra, revealing both spatial and energy bunching.}
	\label{fig:Record2} 
\end{figure*}

We next consider a more realistic scenario using typical parameters of an electron beam generated from an electron gun \cite{LaGrange2025}: a kinetic energy of $W_{0}=79 \  \text{keV}$ and $\Delta W_{0}=4 \  \text{eV}$. 
Using identical laser intensities, we compare the modulation process for different polarization configurations. Table \ref{tab:Formulae} predicts $n_\perp=27$ and $n_\parallel=13$, characteristic times of $T_{s\perp} = 1.73\  \text{ps}$ and $T_{s\parallel} = 0.04\  \text{ps}$, and minimum pulse durations of $\tau_{\min\perp} = 1.6\ \text{as}$ and $\tau_{\min\parallel} = 45\  \text{zs}$ for the orthogonal and parallel-polarized configurations, respectively.
The simulations confirm the predicted pulse numbers $n_\perp$ and $n_\parallel$ [Fig.~\ref{fig:Record2}(b) and (f)] and yield attosecond pulse trains with FWHM durations as short as $1\ \text{as}$ (orthogonal) and $0.14\ \text{as}$ (parallel) [Fig. \ref{fig:Record2} (a) and (e)], consistent with the predicted scaling and demonstrating the strong modulation capability. 

It is worth noting that the energy distribution also exhibits a bunching structure, as indicated by the blue histograms in Fig. \ref{fig:Record2}(d) and (h). This energy modulation accompanies the temporal bunching, with $\gamma_{\max} \approx 15$ under DLA, thereby extending $L_f$ by up to three orders of magnitude compared to the values typical of conventional beams at $\gamma\sim1$. Importantly, the energy spread within each attosecond pulse remains remarkably low---for instance, only $\Delta W / W \approx 0.02\%$ in the orthogonally polarized configuration. 
This combination of low energy spread and high Lorentz factor makes the 1-as electron pulse highly attractive for experimental applications. 

Furthermore, long-term tracking of the pulse trains [Figs. \ref{fig:Record2}(c) and (g)] reveals that the dissipation time is approximately twice the characteristic formation time. Because the energy modulation in the parallel-polarized configuration is several orders of magnitude stronger than in the orthogonal case under identical laser strengths, the attosecond pulse formation proceeds more rapidly in the parallel-polarized configuration, and correspondingly dissipates more quickly.
 
\begin{figure}
   \centering
    \begin{overpic}[width=0.48\textwidth]{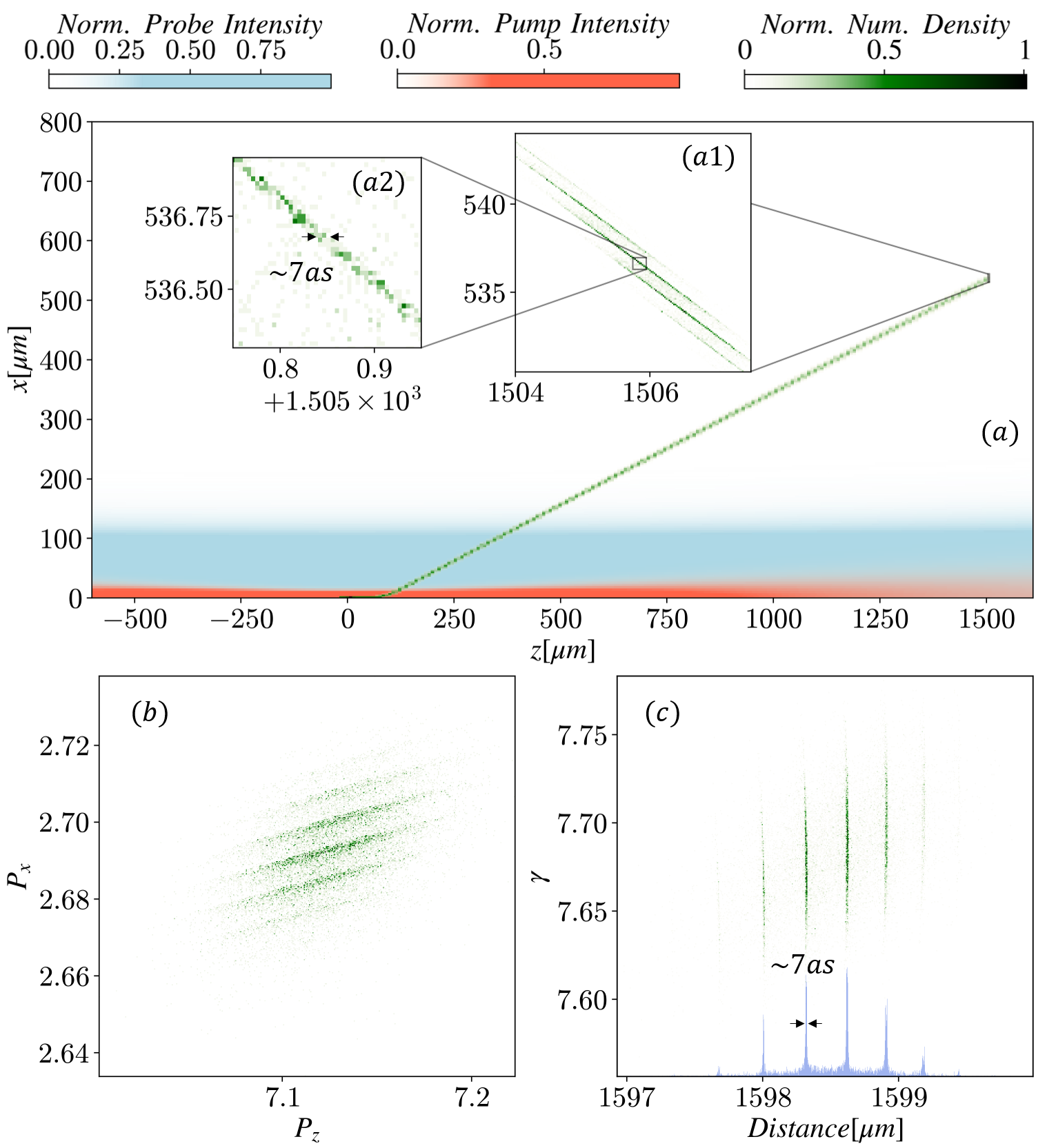}
        \end{overpic}\\

	\caption{Separation of the electron beam via the transverse ponderomotive force. Simulation parameters: $a_{p0}=4.1$, $a_{b0}=0.0001$, $k_p=10k_b=2\pi/(1\ \mu \text{m})$, pump and probe beam waist radii of $10\ \mu \text{m}$ and $100\ \mu \text{m}$. The electron beam with $\beta_0=0.5$ and $\Delta W_{0}=4 \  \text{eV}$ has a waist radius of 10 nm and is initially offset by $0.1\ \mu \text{m}$ along the $x$-axis from the optical axis. The electron beam, pump, and probe durations are 20 fs, 60 fs, and 60 ps, respectively. The pump laser catches up with the electron beam at $z=100\ \mu \text{m}$. (a) Trajectories of pump (red), probe (blue), and electron beam (green); the beam exits the modulation region. (a1),(a2) Magnified views showing a 7 as minimum pulse duration. (b),(c) Momentum and energy distributions of the beam at the detection point. 
 }
	\label{fig:Record3} 
\end{figure}

In practical experiments, the attosecond electron pulse train must be separated from the modulation region of the two lasers for diagnostic purposes. 
We propose to use the transverse ponderomotive force of the pump laser \cite{Bauer1995} to deflect the electron beam away from the modulation region. An illustrative simulation incorporating realistic laser and beam parameters is demonstrated as follows. 

The counter-propagating laser pulses are both x-polarized Gaussian beams with identical divergence angles and different wavelengths. As illustrated in Fig.~\ref{fig:Record3}(a), owing to the transverse ponderomotive force of the pump laser, the electron beam acquires a transverse momentum, resulting in an exit angle of approximately 0.35~rad relative to the optical axis---substantially larger than the laser divergence angle of 0.03~rad.
Notably, although the electron modulation time in this simulation is shorter than $T_s$,  the electron beam exits the laser interaction region with only an energy modulation, which still compresses the beam into an attosecond pulse train during free-space drift. 
The modulated pulse train reaches $\tau_{\min}=7\ \text{as}$ [Fig.~\ref{fig:Record3}(a2)], with a Lorentz factor of 7.7 and $\Delta W/W = 0.6\%$, corresponding to $L_f = 1.6\ \text{mm}$ [Fig.~\ref{fig:Record3}(c)].
Furthermore, as the pump laser continuously overtakes the electron beam, the varying pump laser width produces a position‑dependent ponderomotive force, which introduces small variations in the average energy, energy spread, and exit angle of the individual electron pulses, as shown in Fig.~\ref{fig:Record3}(b) and (c). 
Such differences in exit angles offer a possible route to isolate attosecond pulses via angular separation, offering greater flexibility for experimental manipulation. 

In practice, for low-$\gamma_0$ electron beams, a 100 TW femtosecond pump laser and a 100 kW picosecond probe laser would be sufficient, even with conservative engineering margins.
Given that modern advanced laser facilities already exceed 10 PW in output power \cite{Li2023,Danson2019}, our scheme presents very modest requirements in terms of laser intensity, highlighting its experimental feasibility. 
Beyond its effectiveness for low-$\gamma_0$ electron beams, our analysis demonstrates that the proposed scheme remains viable for beams with higher $\gamma_0$. For instance, using a 1 $\mu\text{m}$ pump laser and a 10 $\text{mm}$ radio-frequency probe wave, the parallel-polarized configuration of our scheme can modulate electron beams with $\gamma_0 \approx 100$. Leveraging its energy bunching capability, our method holds potential for generating X-ray frequency comb \cite{Kalmykov2018}. Moreover, unlike plasma wakefield configurations \cite{Wang2025,xu2022,Wang2026,Angella2026}, our scheme requires at most two laser beams and operates without plasma. This significantly reduces the experimental complexity and improves controllability. 

In summary, a novel scheme is proposed to use two colliding laser fields to accelerate and modulate an electron beam into relativistic few-attosecond pulse trains with tunable periods. The mechanism relies on parametric resonance within the dual-laser system to longitudinally modulate the electron beam, leading to the formation of attosecond pulse trains on a sub-pump-wavelength scale, while DLA simultaneously boosts the pulses to relativistic energies to extend the longitudinal focal length. 
Simulations with realistic sub-relativistic electron beam parameters confirm the generation of electron pulses as short as $1\ \text{as}$ with a maximum Lorentz factor of $15$ and a relative energy spread of $0.02\%$. This work elucidates the ordered modulation regime before stochastic acceleration in dual-laser systems and provides a viable experimental pathway to obtain such high‑quality MeV pulses for attosecond electron-pulse applications.

\vspace{1em}
\begin{acknowledgments}
We are grateful to Professor Chuanxiang Tang and his team at Tsinghua University for their insightful discussions and invaluable inspiration for this work.
This study is supported by the CAS Project for Young Scientists in Basic Research (Grant No. YSBR-141) and the National Natural Science Foundation of China (Grant No. 12375239).
The numerical calculations in this paper were performed on the supercomputing system at the Supercomputing Center of the University of Science and Technology of China. 
\end{acknowledgments}

\bibliography{apssamp}

\newpage
\appendix
\noindent\label{sec:appendixA}\textit{Appendix A: derivation of the Resonance Condition and Related Formulae in orthogonal-polarized configuration}---
For clarity in the following derivation, we set $ c = 1 $, hence $ k_i = \omega_i $ (for $ i = p, b $). The zeroth-order equations of motion are: $P_{px} = a_p, \ P_{pz} = {a_p^2}/{2}, \ \gamma_p = 1 +{a_p^2}/{2}$. The electron position $ z_p $ and time $ t_p $ as functions of the phase variable $ \xi $ are: $z_p=\xi+t_p=-a_{p0}^2[\xi+\sin(2k_p \xi)/(2 k_p)]/4+\mu_0'$. Considering an electron beam with an initial uniform spatial distribution and density $N_e$, the pump laser modulates its longitudinal density profile to: $n(\xi)=N_e\{1+a_{p0}^2[1+\cos(2k_p \xi)]/4\}$. 

The second-order momentum perturbation equation is:
\begin{equation}
\frac{d}{dt} P_{2z} = - \beta_{1y}\partial_z a_b, \gamma_p \beta_{1y} = a_b
\label{eq:app1}
\end{equation}
where $ a_b = a_{b0} \cos[k_b(z + t)] $.
Assuming the electron trajectory is only slightly perturbed by $ \delta z$, and noting $ \mu = z + t = \mu_p + \delta\mu = \mu_p + 2 \delta z $, where $ \mu_p = z_p + t_p $, we expand:
\begin{equation}
\sin(2k_b \mu) \approx\sin(2k_b \mu_p) +4k_b \cos(2k_b \mu_p) \delta z
\end{equation}

Using $ (\gamma_p-P_{pz})dt= -\gamma_p d\xi, \partial_z=d_\xi$ and $ P_{2z} \approx - d_\xi \delta z/2 $, the Eq.(\ref{eq:app1}) becomes:
\begin{equation}
\begin{aligned}
\frac{d^2}{d\xi^2} \delta z - 4a_{b0}^2 k_b^2(d_\xi \mu_p) \cos(2k_b \mu_p) \delta z \\= a_{b0}^2 k_b (d_\xi \mu_p+2d_\xi \delta z) \sin(2k_b \mu_p)
\end{aligned}
\end{equation}
where:
\begin{equation}
\mu_p = - \left( \frac{a_{p0}^2}{2} + 1 \right)\xi - \frac{a_{p0}^2}{4k_p} \sin(2k_p \xi) + 2\mu_0'
\end{equation}
\begin{equation}
\frac{d\mu_p}{d\xi} = - \left( \frac{a_{p0}^2}{2} + 1 \right) - \frac{a_{p0}^2}{2} \cos(2k_p \xi)
\end{equation}

Since modulation relates to the slow-varying component $ \langle \delta z \rangle $, with characteristic time $ T_s \gg 1/\omega_p, 1/\omega_b $, we separate fast and slow variables, yielding:
\begin{equation}
\begin{aligned}
\frac{d^2}{d\xi^2} \langle \delta z \rangle + 4 \left( \frac{a_{p0}^2}{2} + 1 \right) a_{b0}^2 k_b^2 \langle \cos(2k_b \mu_p) \rangle \langle \delta z \rangle\\ = a_{b0}^2 k_b\left[- \left( \frac{a_{p0}^2}{2} + 1 \right) + 2d_\xi\langle \delta z \rangle \right] \langle \sin(2k_b \mu_p) \rangle
\end{aligned}
\label{eq:delta_xi_average}
\end{equation}

Using Bessel function identities:
\begin{equation}
\cos(r \sin\theta) = \sum_{j=0}^\infty \epsilon_{2j} J_{2j}(r) \cos(2j\theta)
\end{equation}
\begin{equation}
\sin(r \sin\theta) = 2 \sum_{j=0}^\infty J_{2j+1}(r) \sin[(2j+1)\theta]
\end{equation}
where $\epsilon_0 = 1, \epsilon_j = 2 \ (j \geq 1)$, we expand:
\begin{equation}
\begin{aligned}
\cos(2k_b \mu_p) = \sum_{j=0}^\infty \epsilon_{2j} J_{2j}\left( \frac{a_{p0}^2 k_b}{2k_p} \right) \cos(4j k_p \xi) \cos( \psi )\\- 2 \sum_{j=0}^\infty J_{2j+1}\left( \frac{a_{p0}^2 k_b}{2k_p} \right) \sin[2(2j+1)k_p \xi] \sin(\psi)
\end{aligned}
\label{eq:cos_expression}
\end{equation}
\begin{equation}
\begin{aligned}
\sin(2k_b \mu_p) = - \sum_{j=0}^\infty \epsilon_{2j} J_{2j}\left( \frac{a_{p0}^2 k_b}{2k_p} \right) \cos(4j k_p \xi) \sin( \psi )\\ - 2 \sum_{j=0}^\infty J_{2j+1}\left( \frac{a_{p0}^2 k_b}{2k_p} \right) \sin[2(2j+1)k_p \xi] \cos( \psi )
\end{aligned}
\label{eq:sin_expression}
\end{equation}
where $\psi=\left( {a_{p0}^2}/{2} + 1 \right) 2k_b \xi - 4k_b \mu_0'$. Non-zero averages $ \langle \cos(2k_b \mu_p) \rangle $ or $ \langle \sin(2k_b \mu_p) \rangle $ occur when $ 2j k_p = ( {a_{p0}^2}/{2} + 1 ) k_b $ or $ (2j+1) k_p = ( {a_{p0}^2}/{2} + 1 ) k_b $.
Let $ n = 2j $ for the even case and $ n = 2j+1 $ for the odd case. $ n $ thus spans the natural numbers. The two cases differ only by a phase shift of $ \pi $. We analyze the even case as an example.

\begin{table}
\caption{\label{tab:table1}%
$ \langle \cos(2k_b \mu_p) \rangle$, $ \langle \sin(2k_b \mu_p) \rangle$ and the state of $\delta z$ with specific values of $4k_b \mu_0'$.
}
\begin{ruledtabular}
\begin{tabular}{cccc}
\textrm{$4k_b \mu_0'$}&
\textrm{$ \langle \cos(2k_b \mu_p) \rangle$}&
\textrm{$ \langle \sin(2k_b \mu_p) \rangle$}&
\textrm{state of $\delta z$}\\
\colrule
$-\pi$&$-J_n\left( \frac{a_{p0}^2 k_b}{2k_p} \right)$&$0$&exponentially grow\\
$-\pi/2$&0&$J_n\left( \frac{a_{p0}^2 k_b}{2k_p} \right)$&increase\\
0&$J_n\left(\frac{a_{p0}^2 k_b}{2k_p}\right)$&$0$&oscillate\\
$\pi/2$&0&$-J_n\left(\frac{a_{p0}^2 k_b}{2k_p}\right)$&decrease\\
$\pi$&$-J_n\left( \frac{a_{p0}^2 k_b}{2k_p} \right)$&$0$&exponentially grow\\
\end{tabular}
\end{ruledtabular}
\end{table}

As shown in Table~\ref{tab:table1}. For $ 4k_b \mu_0' \bmod 2\pi = 0 $, electrons oscillate around a stable point with frequency:
    \begin{equation}
    \widehat{\omega}_{s\perp} =2 \left[ \left( \frac{a_{p0}^2}{2} + 1 \right) a_{b0}^2 k_b^2 J_n\left( \frac{a_{p0}^2 k_b}{2k_p} \right) \right]^{1/2}
    \end{equation}
there are $ n $ such points per half pump wavelength. For $ 4k_b \mu_0' \bmod 2\pi = \pi $ (or $ -\pi $), $ \delta z $ grows exponentially with rate $\widehat{g_s}=\widehat{\omega}_{s\perp} $.
Therefore, we can obtain the oscillation period $ \widehat{T_{s\perp}} \equiv 2\pi/\widehat{\omega}_{s\perp}$. 
Transforming to the laboratory frame leaves the resonance condition unchanged, but $ \widehat{T}_{s\perp} $ dilates to $ \tilde{T}_{s\perp}$:
\begin{equation}
\tilde{T}_{s\perp} = \frac{ \left( \frac{a_{p0}^2}{4} + 1 \right) \pi }{ a_{b0} k_b \left[ \left( \frac{a_{p0}^2}{2} + 1 \right) J_n\left( \frac{a_{p0}^2 k_b}{2k_p} \right) \right]^{1/2}}
\end{equation}

The characteristic modulation time $T_s$, defined as a quarter-period of $\delta z$ closed to the stable fixed point: $T_s\equiv\tilde{T}_s/4$. As a result, electrons initially distributed with different $\mu'_0$ are focused into $n$ distinct attosecond pulses within half a pump wavelength once $t > T_s$.

%
%

\noindent\label{sec:appendixB}\textit{Appendix B: derivation of the Resonance Condition and Related Formulae in parallel-polarized configuration}---
The analysis for the configuration of parallel laser polarizations proceeds similarly to the orthogonal-polarization configuration in many aspects. The first-order equations take the following form:
\begin{equation}
\begin{aligned}
&d_\xi (P_{1x} - a_b) = -P_{1z} \partial_z a_p, 
d_\xi P_{1z} = P_{1x} \partial_z a_p + a_p \partial_z a_b\\
&\Rightarrow d_\xi^2 P_{1z} = \partial_\xi\partial_z (a_p a_b) - P_{1z} \partial_z a_p\\
&\Rightarrow -\frac12 d_\xi^2 \delta z = d_\xi (a_p a_b) - \int P_{1z} \partial_z a_p  d\xi
\end{aligned}
\label{eq:app25}
\end{equation}

The integral term on the right‑hand side contributes to the slow‑varying equation only if $P_{1z}$ contains a component oscillating at the pump frequency. Such a component is absent unless $(a_{p0}^2/2 + 1)k_b = k_p$. For now we exclude this special parameter choice and neglect the integral term in the subsequent discussion.

Using trigonometric identities:
\begin{equation}
\begin{aligned}
a_p a_b = \frac{a_{p0}a_{b0}}{2} \sum_{\sigma=\pm1} \cos\left(k_b\mu + \sigma k_p\xi\right)
      \end{aligned}
      \label{eq:expand1}
\end{equation}
where:
\begin{equation}
\begin{aligned}
&\cos\left(k_b\mu + \sigma k_p\xi\right)\approx \\
 & \cos\left(k_b\mu_p + \sigma k_p\xi\right)
          - 2k_b \sin\left(k_b\mu_p + \sigma k_p\xi\right)\delta z 
\end{aligned}
\label{eq:expand2}
\end{equation}

Substituting Eqs. (\ref{eq:expand1}) and (\ref{eq:expand2}) into Eq.(\ref{eq:app25}) yields:
\begin{equation}
\begin{aligned}
&d_\xi^2\delta z -\\& 4k_ba_{p0}a_{b0} \sum_{\sigma=\pm1}
   \left(k_b d_\xi\mu_p + \sigma k_p\right)
   \cos\left(k_b\mu_p + \sigma k_p\xi\right)\delta z\\
& = a_{p0}a_{b0} \sum_{\sigma=\pm1}
   \left(k_b d_\xi\mu\ + \sigma k_p \right)
   \sin\left(k_b\mu_p + \sigma k_p\xi\right)
   \end{aligned}
\end{equation}

Averaging over the fast oscillations gives the slow‑varying equation:
\begin{equation}
d_\xi^2\langle\delta z\rangle - \widehat{\omega}_{s\parallel}^{2} \langle\delta z\rangle = f
\end{equation}
with:
\begin{equation}
\begin{aligned}
&\widehat{\omega}_{s\parallel}^{2}=\\
& 4k_ba_{p0}a_{b0} \sum_{\sigma=\pm1}
   \Big\langle \left(k_b d_\xi\mu_p + \sigma k_p\right)
   \cos\left(k_b\mu_p + \sigma k_p\xi\right) \Big\rangle
\end{aligned}
\end{equation}
\begin{equation}
\begin{aligned}
f =  \Big\langle a_{p0}a_{b0} \sum_{\sigma=\pm1}
      \left(k_b d_\xi\mu_p + \sigma k_p + 2k_b d_\xi\delta z\right)&\\
      \sin\left(k_b\mu_p + \sigma k_p\xi\right) \Big\rangle&
\end{aligned}
\end{equation}

The essential resonance condition and characteristic time scale are determined by the oscillatory term. Therefore, we omit a detailed discussion of $f$ here.

Expanding $\cos(k_b\mu_p + \sigma k_p\xi)$ using Bessel‑function identities:
\begin{equation}
\begin{aligned}
&\cos\left(k_b\mu_p + \sigma k_p\xi\right)= \\
&\sum_{j=0}^{\infty} \epsilon_{2j}
   J_{2j}\left(\frac{a_{p0}^{2}k_b}{4k_p}\right)
   \cos(4jk_p\xi)
   \cos\left(\frac{\psi}{2} - \sigma k_p\xi\right) \\
& - 2\sum_{j=0}^{\infty}
   J_{2j+1}\left(\frac{a_{p0}^{2}k_b}{4k_p}\right)
   \sin\left[(4j+2)k_p\xi\right]
   \sin\left(\frac{\psi}{2} - \sigma k_p\xi\right)
\end{aligned}
\end{equation}
where $\psi = (a_{p0}^{2}/2 + 1) 2k_b\xi - 4k_b\mu_0'$.

The average $\langle\widehat{\omega}_{s\perp}^{2}\rangle$ becomes non‑zero only when: $4jk_p = \left(a_{p0}^{2}/2 + 1\right)k_b \pm k_p$ or $2(2j+1)k_p = \left(a_{p0}^{2}/2 + 1\right)k_b \pm k_p$, which can be rewritten as:
\begin{equation}
n = \left[\left(a_{p0}^{2}/2 + 1\right)k_b/k_p - 1 \right]/2 \in \mathbb{N}
\end{equation}

The value of $\mu_0'$ influences $\langle\widehat{\omega}_{s\parallel}^{2}\rangle$. For electrons with $2k_b\mu_0' \bmod 2\pi = 0$, the motion consists of oscillations about a fixed point with frequency:
\begin{equation}
\widehat{\omega}_{s\parallel}= 2\left[a_{p0}a_{b0} k_p k_b \sum_{j=0,1} (-)^{j}(n+j)J_{n+j}\left(\frac{a_{p0}^{2}k_b}{4k_p}\right)\right]^{\frac{1}{2}}
\end{equation}

When $n\gg1$, $\widehat{\omega}_{s\parallel}\approx 2\sqrt{a_{p0}a_{b0}k_p k_b n [\mathrm{JJ}]_n(a_{p0}^2k_b/4k_p)}$ , with defining: $[\mathrm{JJ}]_n(\rho)\equiv J_{n}(\rho)-J_{n+1}(\rho)$.
Comparing the distance between adjacent fixed points, we find that the interval in the parallel-polarized configuration is twice that of the orthogonal case, a result also illustrated in Fig.4 of the main text.

Transforming to the laboratory frame, the characteristic modulation formation time becomes:
\begin{equation}
T_{s\parallel} \approx \frac{\left(a_{p0}^{2}/4 + 1\right)\pi}
          {4\sqrt{a_{p0}a_{b0}k_p k_b n [\mathrm{JJ}]_n\left(\frac{a_{p0}^2k_b}{4k_p}\right)}}
\end{equation}
\noindent\label{sec:appendixC}\textit{Appendix C: Discussion of  the energy modulation amplitude and Related Formulae}---
With the discussion in Appendices A and B, the energy amplitude scales as $ \Delta W \propto \Delta P \propto \hat{\omega}_s \Delta \mu_0'$, where $\Delta  \mu_0'$ represents the distance between adjacent fixed points. 
It is $\Delta  \mu_0' = 2\pi /k_b$ for the parallel-polarized configuration, whereas $\Delta  \mu_0' = 2\pi /(2k_b)$ for the orthogonal-polarized configuration. Meanwhile, due to Liouville’s theorem, the minimum pulse duration $\tau_{\min}$ can be obtained as:
\begin{equation}
\tau_{\min}\Delta W=\tau_0 \Delta W_0
\end{equation}
where $\Delta W_0$ represents the initial energy spread and $\tau_0=(1+a_{p0}^2/4)\pi/(nck_p)$.
For an electron beam with a non‑zero initial velocity, the corresponding resonance condition and characteristic time are obtained directly by applying the appropriate Lorentz transformation to the above expressions. 

\end{document}